\documentclass[sigconf,natbib=false]{acmart}
\renewcommand\footnotetextcopyrightpermission[1]{} 
\makeatletter
\renewcommand\@formatdoi[1]{\ignorespaces}
\makeatother
\usepackage{multirow}
\usepackage{tabularx}
\usepackage{longtable}
\usepackage{graphicx}
\usepackage{makecell}
\usepackage{dblfloatfix}
\usepackage[style=ACM-Reference-Format,backend=bibtex,sorting=none]{biblatex}
\AtBeginDocument{%
  \providecommand\BibTeX{{%
    \normalfont B\kern-0.5em{\scshape i\kern-0.25em b}\kern-0.8em\TeX}}}

\setcopyright{none}
\begin{document}

\title{Development Frameworks for Microservice-based Applications: Evaluation and Comparison}

\author{Hai Dinh-Tuan}
\email{hai.dinhtuan@tu-berlin.de}
\affiliation{%
  \institution{Technische Universität Berlin}
  \streetaddress{Ernst-Reuter-Platz 7}
  \city{Berlin}
  \country{Germany}
}

\author{Maria Mora-Martinez}
\email{m.moramartinez@tu-berlin.de}
\affiliation{%
  \institution{Technische Universität Berlin}
  \streetaddress{Ernst-Reuter-Platz 7}
  \city{Berlin}
  \country{Germany}}

\author{Felix Beierle}
\email{beierle@tu-berlin.de}
\affiliation{%
  \institution{Technische Universität Berlin}
  \streetaddress{Ernst-Reuter-Platz 7}
  \city{Berlin}
  \country{Germany}}

\author{Sandro Rodriguez Garzon}
\email{sandro.rodriguezgarzon@tu-berlin.de}
\affiliation{%
  \institution{Technische Universität Berlin}
  \streetaddress{Ernst-Reuter-Platz 7}
  \city{Berlin}
  \country{Germany}}

\renewcommand{\shortauthors}{Dinh-Tuan and Mora-Martinez, et al.}

\begin{abstract}
The microservice architectural style has gained much attention from both academia and industry recently as a novel way to design, develop, and deploy cloud-native applications.This concept encourages the decomposition of a monolith into multiple independently deployable units. A typical microservices-based application is formed of two service types: \textit{functional services}, which provide the core business logic, and \textit{infrastructure services}, which provide essential functionalities for a microservices ecosystem. To improve developers' productivity, many software frameworks have been developed to provide those reusable infrastructure services, allowing programmers to focus on implementing microservices in arbitrary ways. In this work, we made use of four open source frameworks to develop a cloud-based application in order to compare and evaluate their usability and practicability. While all selected frameworks promote asynchronous microservice design in general, there are differences in the ways each implements services. This leads to interoperability issues, such as  message topic naming convention. Additionally, a key finding is the long startup times of JVM-based services that might reduce application's resiliency and portability. Some other advantages come directly from the programming language, such as the ability of Go to generate native binary executables, which results in very small and compact Docker images (up to 78\% smaller compared to other languages).
\end{abstract}



\maketitle
\pagestyle{plain}

\section{Introduction}
The microservice architectural style emerged from the \textit{Service-oriented architecture} and the \textit{Domain-Driven Design} architectural patterns with a strong focus on DevOps practices \cite{lewis2014microservices}. Not only has it been successfully adopted in multimillion-user companies such as Netflix or Amazon, but also received significant attention from the academia as a robust and scalable method to build cloud applications. In short, the microservice architecture advocates a new software design, in which the application is structured into services. Theses services are independently developed, updated, and deployed, thus significantly improving the flexibility of the application. Similarly, the overall scalability and fault tolerance of the application are enhanced. However, these advantages also come at a cost. Applications get more complex and latency increases because the communication channel is more susceptible to network delays. 

To reduce the burdens created by distributed systems' complexities, several frameworks have been developed specifically for microservices. They provide boilerplate code and abstractions to allow the developers to focus on the business logic. By exploring implementation options for an example application, this paper proposes three criteria for evaluating and comparing frameworks. For \textit{features}, we compare frameworks across three main categories, which are maturity, development supports, and built-in features. For \textit{design patterns}, we categorized supported ones into six groups. Lastly, for \textit{performance}, we relied on two main criteria, which are end-to-end latency performance and resource consumption. This paper therefore delivers a three-fold contribution:

\begin{enumerate}
  \item A review of different approaches to implement microservices, which is exhibited in four microservice frameworks selected.
  \item A comparison model for microservices frameworks, based on three main criteria: features, design patterns, and performance.
  \item The implications of adopting a framework in developing microservices.
\end{enumerate}

In the remainder of this paper, Section \ref{related} gives a brief account of the evaluations and comparisons have been done so far for microservices. Then, we introduce the motivating use case and the architecture for a toll system's prototype in Section \ref{usecase}. Section \ref{features} presents a feature-wise comparison while Section \ref{performance} shows practical performance results. We discuss the results in Section \ref{discussion}, draw the overall conclusion in Section \ref{conclusion}, and list future research directions in Section \ref{future}.

\section{Related work} \label{related}
One of the most well known definitions of software framework proposed by Johnson describes it as  "a reusable design of all or part of a system that is represented by a set of abstract classes and the way their instances interact" or "the skeleton of an application that can be customized by an application developer" \cite{johnson1997components}. With a software framework, developers are able to reuse both design and code to focus on solving business specific needs \cite{johnson1988designing}. Therefore, framework-based software development can be seen as an activity to modify and extend the framework \cite{morisio2002quality}. In addition, a framework can improve programmer productivity and enhance the quality, performance, reliability and interoperability of software \cite{fayad1997object, mattsson1999effort, johnson1988designing}. However, a software framework can also lead to prolonged development time, as developers have to learn and get familiar with the framework \cite{srinivasan1999design}. Also, adopting a framework might result in more complicated code structure and larger software footprints. 

As frameworks are usually designed to solve a narrow set of problems by targeting a certain type of applications \cite{fayad2000enterprise}, there are more and more frameworks specialized to ease, speed up and improve development of applications with a  microservice-based architectural design. There have been several works study the performance of microservices. \cite{ueda2016workload, amaral2015performance}. Others propose criteria to evaluate microservices such as \cite{aderaldo2017benchmark, sriraman2018mu}. However, to the best of our knowledge, there is no research article that compares the performance and approaches by using different programming languages and frameworks to build the same application. This paper addresses this exact gap to compare and evaluate different framework-based development approaches for microservices. 

\section{Motivating use case} \label{usecase}
One of the biggest issues in urban areas is the decline of air quality, which causes 4.2 million premature deaths every year, according to a recent report by the World Health Organization (WHO) \cite{who2019}. In the same report, human mobility activities has been identified as the major sources of outdoor air pollution, including fossil fuels combustion from motor vehicles. 

In many countries, policy-makers have introduced various regulations to address these problems, one of them is to implement a pollution toll. However, current toll schemes charge fixed fees based on static criteria such as vehicle type or toll area \cite{wang2017study}. In this work, inspired by \cite{garzon2019pay}, we propose a novel fine-grained system to calculate toll fee proportionally to the pollution level of each area. This means the more polluted an area is, the higher fee per driven kilometre is applied for all vehicle in that area. One obvious advantage of this approach is that this system allows a dynamic toll rate based on real-time air pollution level. In addition, this is very scalable, as it only uses location data of the vehicle in short intervals to calculate the toll fee. The fairness of this system can be further improved by adding information about the vehicle's model and speed (estimated from the location data).

\begin{figure}
	\includegraphics[width=\linewidth]{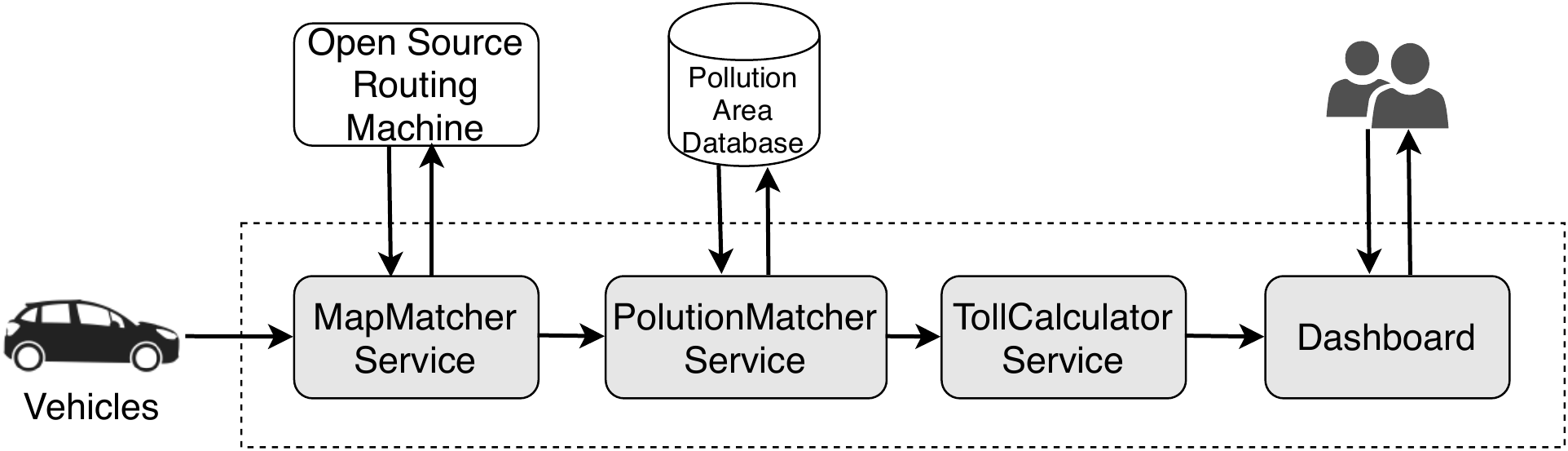}
	\caption{Overview architecture of the proposed air pollution-aware toll system.}
	\label{fig:overview}
\end{figure}

The hypothetical toll system adopts the event-driven microservice architecture, in which each microservice is responsible for a specific task and data is exchanged as JSONs through a message broker using the publish-subscribe asynchronous messaging pattern. Fig. \ref{fig:overview} illustrates the architecture, in which the arrows among microservices are not direct communication but rather an indirect communication via a message broker. The processing pipeline is explained in detail below:

\begin{itemize}
	\item The vehicles update their GPS coordinates information by sending JSON-based messages via the message broker.
	\item The \textit{Map Matcher Service} subscribes to \textit{Location update messages} from the vehicles. Then it uses the APIs provided by a local instance of OSRM (Open Source Routing Machine) \cite{huber2016calculate} to map the reported coordinates to a route on streets. Then the driven route will be sent to the message broker in the \textit{Route Messages}.
	\item The \textit{Pollution Matcher Service} receives the \textit{Route messages} and compares them with an air pollution database. In this PostGIS database, polluted areas are stored as polygons with corresponding pollution levels. The received \textit{route} is matched against these polygons and in case the route crosses multiple areas with different pollution levels, it will be broken down into several \textit{segments}. These segments together with corresponding pollution levels are sent by the \textit{Pollution Matcher Service} to the message broker via \textit{Segment Messages}.
	\item The \textit{Toll Calculator Service} calculates the fare for each vehicle based on the \textit{Segment Messages}. It calculates the travel distance of each segment and, based on the pollution level, returns the accumulative toll fee for each vehicle. Those values are sent to the message broker as \textit{Toll Messages}.
	\item Upon receipt of the \textit{Toll Messages}, the \textit{Dashboard} displays the cumulative toll fee for each car and visualizes each vehicle's route on an interactive map.
\end{itemize}

\section{Frameworks} \label{frameworks} 
There is a large availability of frameworks for different programming languages. For the implementation of the toll system, we selected four different microservices frameworks: Go Micro (Go), Moleculer (JavaScript with Node.js), Lagom (Scala), and Spring Boot/Spring Cloud (Java). They are chosen to reflect the variety of tools that can be used to develop microservices. With respect to programming languages, Java is consider an “old" language while Scala and Go both have been introduced in the last decade, and Node.js is a new run-time environment for JavaScript. In terms of framework design philosophy, although Spring Boot is widely used for microservices, it is not designed specifically for such applications like the other three. Each framework provides a different approach to build microservice-based applications. In this section, we briefly discuss the main differences.

\paragraph{Go Micro} provides an abstraction for distributed systems development using Remote Procedure Call (RPC) and Event-driven communication written in Go. This opinionated framework is provided as a set of pluggable building blocks that can be included or excluded in a project depending on the application's specific requirements. This design reduces the complexity in development. They cover a wide range of functionalities, from sync/async messaging, message encoding, service discovery, to dynamic configuration. To define a microservice with Go Micro, the framework emcourages developers to firstly define the interfaces (default by ProtoBuf), and then implement them. Afterwards, the service can be initialized and registered with the built-in service registry.

\paragraph{Lagom} is another opinionated microservice framework to build reactive microservices in Scala or Java. Developed and maintained by the same company behind Akka and Play Framework, Lagom is built to be easily integrated into these two technologies. Lagom supports and encourages using \textit{Event sourcing} and \textit{Command Query Responsibility Segregation (CQRS)} for data persistence. Lagom recommends developers to separate the \textit{service interface} and \textit{service implementation}, which are organized as sub-projects within the same build. While the former contains the interface descriptions, along with data models, the latter implements the service logic.

\paragraph{Moleculer} is a microservice framework for JavaScript using Node.js runtime built on Chrome's V8 JavaScript engine. The main design philosophy of this framework is to produce low latency event-driven applications. Moleculer also provides pluggable modules for various functions such as gateways, database, serialization, etc. The framework consists of three main components: the \textit{transporter}, the \textit{service} and the \textit{service broker}. While the \textit{service} defines the general structure for each microservice, the \textit{transporter} is an abstraction for the communication bus, and the \textit{service broker} binds \textit{service} with \textit{transporter}.

\paragraph{Spring Boot \& Spring Cloud} are two projects from the Spring ecosystem. While Spring Boot is an opinionated way to build stand-alone Spring applications, Spring Cloud provides a set of tools to quickly build several common patterns in distributed systems such as service discovery, configuration management, etc. That means Spring Boot is not specifically designed for developing microservices but rather a full stack development environment. Spring Boot has to rely on Spring Cloud for microservices-specific capabilities.

\begin{table*}[!b]
\caption{Features of microservices frameworks}
\begin{tabularx}{\linewidth}{lXXXXXX}
\toprule
 &
   &
   &
  Spring Boot &
  Lagom &
  Moleculer &
  Go Micro \\ \hline
{\multirow{7}{*}{\rotatebox[origin=c]{90}{Maturity}}} &
  \multirow{6}{*}{GitHub stats} &
  Initial release date &
  July 10th 2014 &
  March  3rd 2016 &
  February 16th 2017 &
  December 6th 2017 \\ \cline{3-7} 
 &
   &
  Releases &
  146 &
  76 &
  86 &
  70 \\ \cline{3-7} 
 &
   &
  Commits &
  25357 &
  2397 &
  3222 &
  2700 \\ \cline{3-7} 
 &
   &
  Watch &
  3348 &
  165 &
  123 &
  452 \\ \cline{3-7} 
 &
   &
  Star &
  45626 &
  2370 &
  3322 &
  11748 \\ \cline{3-7} 
 &
   &
  Contributors &
  656 &
  126 &
  65 &
  77 \\ \hline
{\multirow{13}{*}{\rotatebox[origin=c]{90}{Development supports}}} &
  \multirow{2}{*}{Documentation} &
  Documentation &
  Very Extensive &
  Extensive &
  Less extensive &
  Limited \\ \cline{3-7} 
 &
   &
  Sample code &
  Yes &
  Yes &
  Yes &
  Yes \\ \cline{2-7} 
 &
  Supported languages &
  Supported languages &
  Java, Kotlin, Groovy, Beanshell &
  Java, Scala &
  Javascript/NodeJs &
  Go \\ \cline{2-7} 
 &
  \multirow{4}{*}{Community supports} &
  Stack OverFlow tags &
  162688 &
  296 &
  38 &
  29 \\ \cline{3-7} 
 &
   &
  Chat platform &
  Gitter &
  Gitter &
  Gitter &
  Slack \\ \cline{3-7} 
 &
   &
  Forum &
  Yes &
  Yes &
  Yes &
  Yes \\ \cline{3-7} 
 &
   &
  GitHub &
  Yes &
  Yes &
  Yes &
  Yes \\ \cline{2-7} 
 &
  \multirow{2}{*}{Commercial supports} &
  Training course &
  Yes &
  Yes &
  No &
  No \\ \cline{3-7} 
 &
   &
  Certification &
  Yes &
  Yes &
  No &
  No \\ \cline{2-7} 
 &
  \multirow{4}{=}{Development environment} &
  Build tool \& Dependency Management &
  Maven (default), Gradle, Ant &
  sbt &
  Default to npm &
  Default to Go Build \\ \cline{3-7} 
 &
   &
  Hot reload &
  Yes, but not by default &
  Yes, supported by default with sbt &
  Yes, enabled by default &
  N/A \\ \cline{3-7} 
 &
   &
  Template generator &
  Spring Initialzr (web-based) &
  Lagom Tech Hub  Project Starter (web-based) &
  moleculer-cli command &
  micro new command \\ \cline{3-7} 
 &
   &
  Interactive development console &
  N/A &
  sbt development console, only for development purposes &
  REPL console: list nodes, services, actions, events, environment variables, etc. &
  Interactive Micro CLI, to manage the services \\ \hline
{\multirow{10}{*}{\rotatebox[origin=c]{90}{Built-in features}}} &
  \multirow{5}{*}{Essential services} &
  API Gateway &
  Yes &
  Yes &
  Yes &
  Yes \\ \cline{3-7} 
 &
   &
  Service discovery &
  Yes &
  Yes &
  Yes &
  Yes \\ \cline{3-7} 
 &
   &
  Service registry &
  Yes &
  Yes &
  Yes &
  Yes \\ \cline{3-7} 
 &
   &
  Load balancing &
  Spring Cloud Load Balancer &
  Relies on underlying Akka &
  server-side load balancing by default &
  proxy package \\ \cline{3-7} 
 &
   &
  Serialization &
  JSON by Jackson &
  JSON (defualt), ProtoBuf, Apache Avro, Kryo &
  JSON (default), Avro, MsgPack, Notepack, ProtoBuf, Thrift, and custom modules &
  JSON or ProtoBuf \\ \cline{2-7} 
 &
  \multirow{4}{*}{Other features} &
  Integration with Slack &
  N/A &
  N/A &
  N/A &
  bot package \\ \cline{3-7} 
 &
   &
  Interactive CLI & 
  Spring Boot CLI &
  sbt &
  moleculer REPL &
  Micro CLI package  \\ \cline{3-7} 
 &
   &
  Web dashboard &
  N/A &
  N/A &
  N/A &
  web package \\ \cline{3-7} 
 &
   &
  Caching &
  JCache,EhCache 2.x, Hazelcast, Infinispan, Couchbase, Redis, Caffeine, etc. &
  N/A &
  built-in cache, heap memory, LRU cache module, Redis-based, or custom cacher &
  N/A \\ \bottomrule
\end{tabularx}\\ 
\label{tab:feature-table}
\end{table*}

\section{Feature comparison} \label{features}
In this section, we provide an extensive evaluation of those frameworks with respect to 2 main criteria: feature \textit{features} and \textit{design patterns}. For the former, we further sub-categorize it into (1) Maturity, (2) Supports for microservice development, and (3) Built-in features. All of these details will be summarized in Table \ref{tab:feature-table} and Table \ref{tab:design-table}. Note that, a "N/A" does not necessary mean that the framework lacks a certain feature, but rather that the feature requires external libraries, which has not yet been officially integrated into the framework.

\subsection{Main features and design philosophy}
All four microservice frameworks focus on providing infrastructure services for microservice-based applications. In fact, the separation between infrastructure ser vices and functional services has been well discussed in the literature \cite{dinh2019maia}. While infrastructure services provide essential capability such as service registry or service discovery, functional services are the ones that provide actual business logic. 

As observed in the four implementations, there are two main approaches from the different frameworks: they can either implement their own services or provide integration with existing solutions. In addition, as open source projects, besides the official code, they can provide community-maintained plugins and modules such as in Go Micro and Moleculer.

\subsection{Maturity}
To quantify the maturity of a open sourced software framework, we take a look at the release cycle and metrics from open source websites for code hosting such as GitHub as used in previous research \cite{mcdonald2013performance, dozmorov2018github}. 

Given that Spring Boot has been initially releases in 2014, while Lagom's first release was in 2016 and both Go Micro and Moleculer in 2017, Spring Boot has become one of the most mature frameworks. That also explains how the amount of releases from Spring outperforms the others twice, as seen in Table \ref{tab:feature-table}. However, as Spring Boot is also a popular framework to build generic Java applications, this might not fully reflect the maturity of Spring in developing microservices specifically. 

\subsection{Supports for microservice development}
Given the difference in maturity among the selected frameworks, one can expect different levels of support and learning curves. Indeed, the documentation of Spring Boot is very extensive compared to the others. With a huge community and long existence on the market, it is easy to find support from developer communities such as StackOverflow. Besides, Spring and Lagom are backed by Pivotal and Lightbend respectively, so they can offer commercial support for large organizations, while Moleculer and Go Micro must rely on community-based supports.

Build tools and dependency management solution plays an important roles to significantly improve the productivity. Boilerplate codes for new microservices can be quickly generated following certain templates, however each framwework supports different level of customization. For example, Lagom and Moleculer only allow to set up project name and service name, while the other two allow specifying dependency packages. In addition, features such a hot reloading or interactive development console can further shorten development cycles, as they help developers to immediately see the effects of new code at runtime.

\subsection{Design patterns}
To achieve robustness in microservices, there are certain recommendations to follow, such as minimizing service dependencies or separating data storage \cite{pahl2016microservices}. These non-functional properties materialize in different design patterns. To understand how each selected framework design philosophy differs from each other, we briefly discuss them in this section and provide a summary in Table \ref{tab:design-table}. 

To provide API access for external users, all four frameworks support the \textit{API Gateway} pattern. In this design pattern accesses to individual microservices are not directly provided to the user. Instead, a gateway is used as a single entry point for the application, exposing APIs, routing messages to responsible microservices, or even translating messaging between different protocols, such as in Go Micro. 

Another design pattern advocated by all frameworks is the \textit{Database per service} design pattern. Microservice design helps decoupling services as they can be developed, deployed, and updated without breaking other services. However, data can create hidden dependencies among services. Therefore,  each microservice should manage its own data and no service can modify data belonging to another service. In contrast, shared databases among multiple services is not encouraged by any framework. A more advanced pattern, such as Command-Query Responsibility Segregation with Event Sourcing, is even part of the framework philosophy like in the case of Lagom.

In a environment with multiple services being independently deployed, it is crucial for the operators to have an overview of the application and insights into underlying operations \cite{bakshi2017microservices}. Therefore, all four frameworks provide extensive integration with different logging tools. However, when it comes to metrics collecting, while Moleculer has a built-in module collecting and publishing metrics to various outputs, Spring Boot and Go Micro provide a metrics collector as a pluggable module; Lagom has to rely on Lightbend cloud platform for those capabilities. This is because Lagom has been designed to simplifying microservices application development and deployment on Lightbend's current offer in cloud platform.

Similarly, a technique used for monitoring and analyzing distributed systems called \textit{Distributed Tracing} is also supported by all frameworks. In this technique, as a request travels through the system, a \textit{unique identifier} is added, allowing details about the path, time spend spent in each services to be centrally captured and visualized. Although no framework provides its own implementation of a monitoring service supporting Distributed Tracing, Spring Boot, Moleculer, and Go Micro are fully integrated with external solutions such as Prometheus. In contrast, Lagom microservices have to rely again on Lightbend's platform.

With the recent growth of container-based cloud platforms, Docker and Docker-based orchestration platforms like Kubernetes, Mesosphere DC/OS, OpenShift, etc., have become the de facto standard. All of the frameworks have supports for Docker deployment, except Moleculer's support for Kubernetes still missing. Both Spring Boot and Lagom can build executables that can be deployed on Heroku, Azure, etc.

Regarding the communication patterns among microservices, all of the frameworks encourage using asynchronous messaging as the main way for exchanging information. Asynchronous messaging can decouple services, minimizing dependencies, and improve the fault tolerance of the application as a whole \cite{richter2017highly}. In fact, being able to operate asynchronously is one of the design criteria to get the most out of the microservice architectural style \cite{garriga2017towards}. While most of the widely used protocols such as AMQP, MQTT, are fully supported by all, each framework implements the communication channel differently: Spring, Lagom, and Go Micro provides different integration modules for message brokers, while Moleculer differentiates two communication types. \textit{Local services} use an in-memory local bus for exchanging messages. Remote services relies on an abstraction layer called \textit{Transporter}, which can deliver events, requests, and responses. By default, NATS is used as the broker, however, other implementations cannot be added directly, because Moleculer automatically adds prefixes to messages topics.

In a microservices environment with a large number of distributed services, it is utmost crucial to prevent cascading failures. All four frameworks support the \textit{Circuit Breaker} pattern. In this design, in case of failure, requests are rejected immediately, thus avoiding errors.

\begin{table*}[!b]
\caption{Design patterns of microservice frameworks}
\begin{tabularx}{\linewidth}{lXXXXX}
\toprule
 &
   &
  Spring &
  Lagom &
  Moleculer &
  Go Micro \\ \hline
{{\rotatebox[origin=c]{90}{Integration}}} &
  API Gateway &
  Spring Cloud Gateway, built on top of Spring MVC, Netty &
  Default embedded gateway based on Akka HTTP &
  moleculer-web for request routing &
  Go API libraries for protocol translation and request routing \\ \hline
{\multirow{4}{*}{\rotatebox[origin=c]{90}{Database}}} &

  Database per service &
  Supported &
  Cassandra, MySQL, PostgreSQL, Oracle, Couchbase, Microsoft SQL, H2 &
  MongoDB, MySQL, PostgreSQL, SQLite, MSSQL. Default to CRUD actions &
  Only provides a Store interface for additional storage in the future \\ \cline{2-6} 
 &
  Shared database &
  Supported &
  Discouraged &
  Discouraged &
  N/A \\ \cline{2-6} 
 &
  CQRS/Event Sourcing &
  Supported &
  Supported by default &
  N/A &
  N/A \\ \hline
{\multirow{4}{*}{\rotatebox[origin=c]{90}{Observability}}} &
  Log Aggregation &
  Logback(default), Java Util Logging, Log4J2 &
  SLF4J based on Logback (default), Log4J2 &
  Console output, Pino, Bunyan, Winston, Log4js, Datadog &
  Logrus, Zap, Zerolog \\ \cline{2-6} 
 &
  Performance Metrics &
  Prometheus, Datadog, Netflix Atlas &
  Provided by Lightbend platform &
  Prometheus, StatsD, Datadog &
  Prometheus. \\ \cline{2-6} 
 &
  Distributed Tracing &
  Supported with Spring Cloud Sleuth &
  Provided by Lightbend, which supports OpenTracing &
  Built-in tracing module works with Zipkin, Jaeger, Datadog &
  Amazon AWS X-Ray, UUID plugin to trace messages manually \\ \cline{2-6} 
 &
  Health Check &
  supported with Spring Actuator &
  N/A &
  Built-in &
  sidecar implementation via RPC \\ \hline
{\multirow{5}{*}{\rotatebox[origin=c]{90}{Cross-cutting concerns}}} &

  External configuration &
  Spring Cloud Netflix for client and server-side, Netflix Archaius &
  N/A &
  External configuration files can be loaded via moleculer-runner &
  Supported via Go Config \\ \cline{2-6} 
 &
  Service  Registry &
  Netflix Eureka. Also Kubernetes, Consul, Apache Zookeeper, Alibaba Nacos &
  Provided by default &
  Built-in service registry &
  Built-in, can be accessed via a Web Dashboard \\ \cline{2-6} 
 &
  Client-side service discovery &
  Client-side load balancing supported with Ribbon &
  Provided in the ServiceLocator &
  N/A &
  mdns, etcd \\ \cline{2-6} 
 &
  Server-side service discovery &
  N/A &
  Supported &
  Default &
  mdns, etcd \\ \cline{2-6} 
 &
  Deployment platform &
  Docker, Kubernetes, Cloud Foundry, Azure, Heroku, Amazon AWS, OpenShift, Boxfuse &
  Docker, Kubernetes, OpenShift, MesosphereDC/OS &
  Docker/Docker Compose, Kubernetes under development &
  Local, Docker, Kubernetes \\ \hline
{\multirow{3}{*}{\rotatebox[origin=c]{90}{Communication}}} &

  Synchronous messaging &
  Supported &
  Supported &
  Supported &
  Supported \\
   \cline{2-6} 
 &
  Asynchronous messaging &
  RabbitMQ, Apache Kafka, Amazon Kinesis, Google PubSub &
  Message Broker API for Kafka &
  Preferred: NATS, Kafka, Redis, MQTT, AMQP &
  PubSub is preferred. Default embedded NATS server \\ \cline{2-6} 
 &
  Remote Procedure Invocation &
  RMI, Spring's HTTP Invoker, Hessian, Burlap &
  Akka gRPC library HTTP/JSON &
  RPC is implemented by default &
  Supports gRPC. \\ \hline
{\multirow{7}{*}{\rotatebox[origin=c]{90}{Fault Tolerance}}} &

  Circuit Breaker &
  Supported with Netflix Hystrix &
  Used for all service calls by default &
  Built-in, threshold-based &
  Plugins. \\ \cline{2-6} 
 &
  Retry &
  Spring-retry, Retryable annotation &
  N/A &
  Exponential back-off retry &
  Resend the requests to another node \\ \cline{2-6} 
 &
  Timeout &
  For HTTP components &
  Circuit breaker: call and reset timeouts &
  Set timeout values for requests &
  N/A \\ \hline
\end{tabularx} \\
\label{tab:design-table}
\end{table*}

\section{Performance evaluation} \label{performance}
\subsection{Test setup}
We implemented the pipeline as described in section \ref{usecase} with each framework. For the evaluation test, the following versions of the tools and frameworks have been used: Spring Boot 2.2.4 (Java 1.8), Go Micro 1.18.0, (Go 1.13), Moleculer 0.13.0 (Node.js 13.1.4), Lagom 1.6.0 (Scala 2.12.8), NATS 2.1.2. To ensure the consistency of input data across tests, we built a \textit{Traffic Simulator} to simulate vehicles. This component is based on the \textit{Simulation of Urban MObility (SUMO)} \cite{behrisch2011sumo}, and provides a traffic simulation in a European city using data from the \textit{OpenStreetMap}. From the \textit{Dashboard}, the simulation can be configured by the number of simulated vehicles and update frequency. The results presented in this paper is from simulations with 10 vehicles, and they update their location every 5 seconds.

The machine where the tests were carried out has an Intel Core i5 processor and 16GB of RAM. All microservices are containerized using Docker and deployed using Docker Compose, as this is a popular choice to achieve platform-independent deployment strategy \cite{ueda2016workload}. We have considered latency and resource consumption as the key criteria to measure the performance of the different frameworks. Another resource metric is the footprint of the application, given by the size of the Docker images.

The selected usecase represents some typical microservices types, which should be taken into account when application being evaluated. The \textit{Map Matcher} portrays a service which consumes external API, the \textit{Pollution Matcher} constantly exchanges data with an external database, while the \textit{Toll Calculator} performs all the calculation internally.

\subsection{Slow start of JVM-based implementations} 
\begin{figure}
	\includegraphics[width=\linewidth]{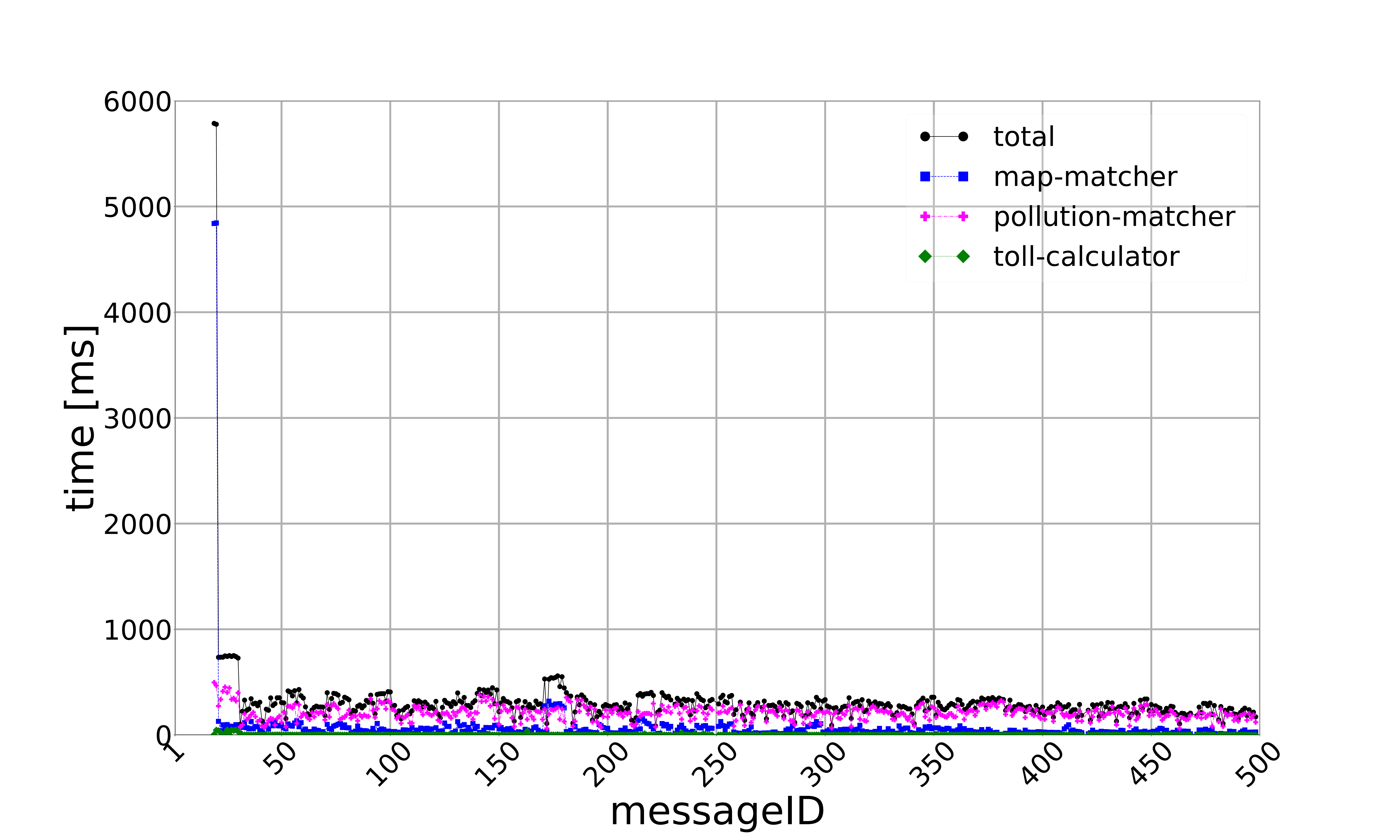}
	\caption{Long latency for first messages in Lagom.}
	\label{fig:lagom_long_latency}
\end{figure}

We noticed long latencies in the first messages processed in Spring Boot and Lagom implementations compared to subsequent requests as illustrated in Fig. \ref{fig:lagom_long_latency}.

The reason for such a strange behavior, which is not found in other implementations, lies in the Java Virtual Machine (JVM) warm up phase \cite{lion2016don}. In this test, to reduce the effects of the warm up phase, we only evaluate the framework performance after five message iterations, i.e. skipping the first 50 messages.

\subsection{End-to-end latency performance}
\begin{figure}
	\includegraphics[width=\linewidth]{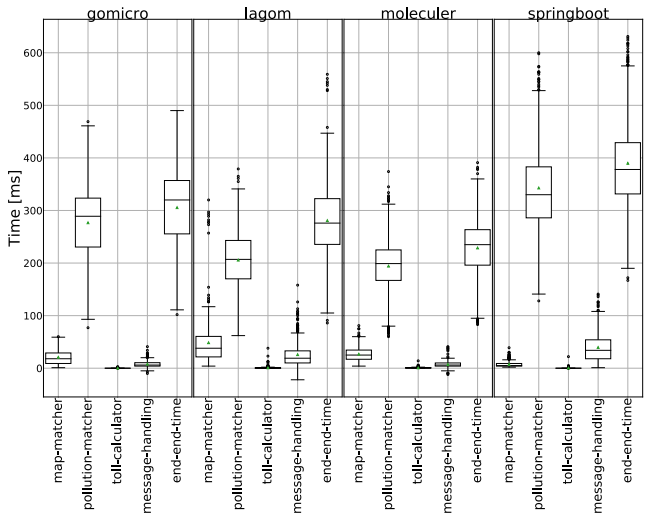}
	\caption{Comparison of latency performance.}
	\label{fig:toll_performance}
\end{figure}
Fig. \ref{fig:toll_performance} summarizes the end-to-end latency of each framework. All four implementations show similar behavior in regard to latency, although to different extents. There are still some noticeable patterns shared by all implementation. For example, the slowest microservice is always the \textit{Pollution Matcher} that has to access the PostGIS database, it accounts from 73.27\% in Lagom to 90.59\% of the total time in Go Micro implementation. By contrast the \textit{Toll Calculator} has a minimal contribution to the total end-to-end time since it can perform all the processes internally. From the results, it is clear that database access is very slow compared to other I/O bound tasks. This illustrates difficulties in managing service's state, in which the they should be externalized into a database to create totally stateless services. In many designs, the microservices are forced to follow the 12-factor app principle, i.e. stateless, but generally it is either very difficult or not optimized to build applications solely from stateless microservices.

Communication overhead is more dispersed for Lagom and Spring, while Go Micro and Moleculer have a more consistent delay. Even with slower communication, Lagom performs in general better than Go Micro. 

\subsection{Resource consumption} 
\begin{figure}
	\includegraphics[width=\linewidth]{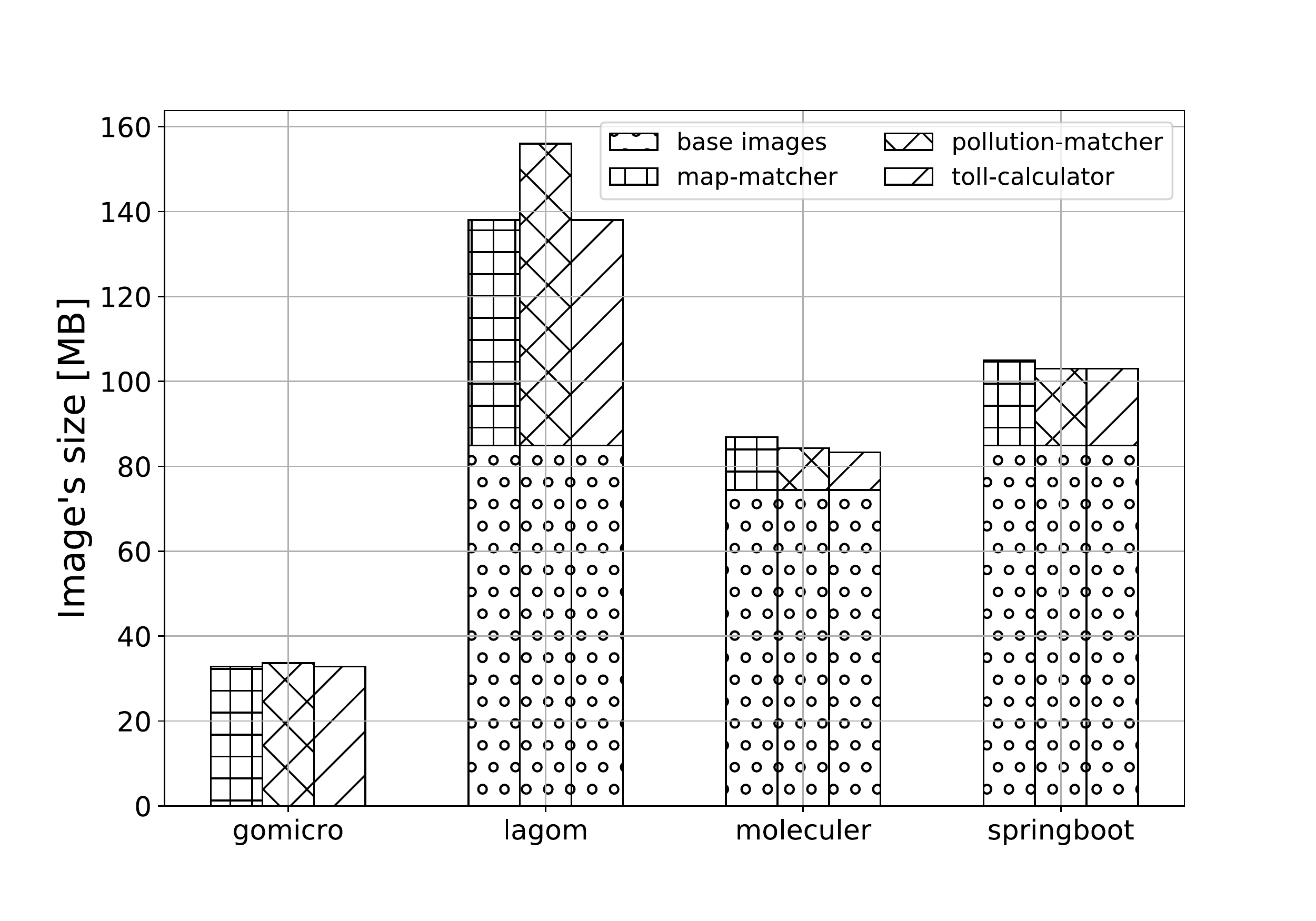}
	\caption{Docker images size comparison.}
	\label{fig:docker_image_size}
\end{figure}

\paragraph{Docker images size} We evaluated the Docker image size produced by each framework, as smaller images can be deployed and created quickly, leading to faster update cycles \cite{kim2018gpu}. In this setup, we used Alpine-based images: Spring uses openjdk:8-jre-alpine, Lagom uses openjdk:8-jre-alpine, Moleculer uses mhart\slash alpine-node:10, and Go Micro uses \textit{golang} base image as build image and \textit{scratch} as deployment container. The results are summarized in Fig. \ref{fig:docker_image_size}. As Go applications can be built as binary files, they can use \textit{scratch} - a minimal empty Docker image. Therefore services developed with this language are very compact compared to others, for example, the \textit{Pollution Matcher} written in Go Micro is 78\% smaller the implementation with Lagom (33.6 MB vs. 156 MB). Moleculer also has the ability to create smaller Dockerized services compared to Lagom or Spring Boot, as these two frameworks require a JVM to run the source code.

\begin{figure}
	\includegraphics[width=\linewidth]{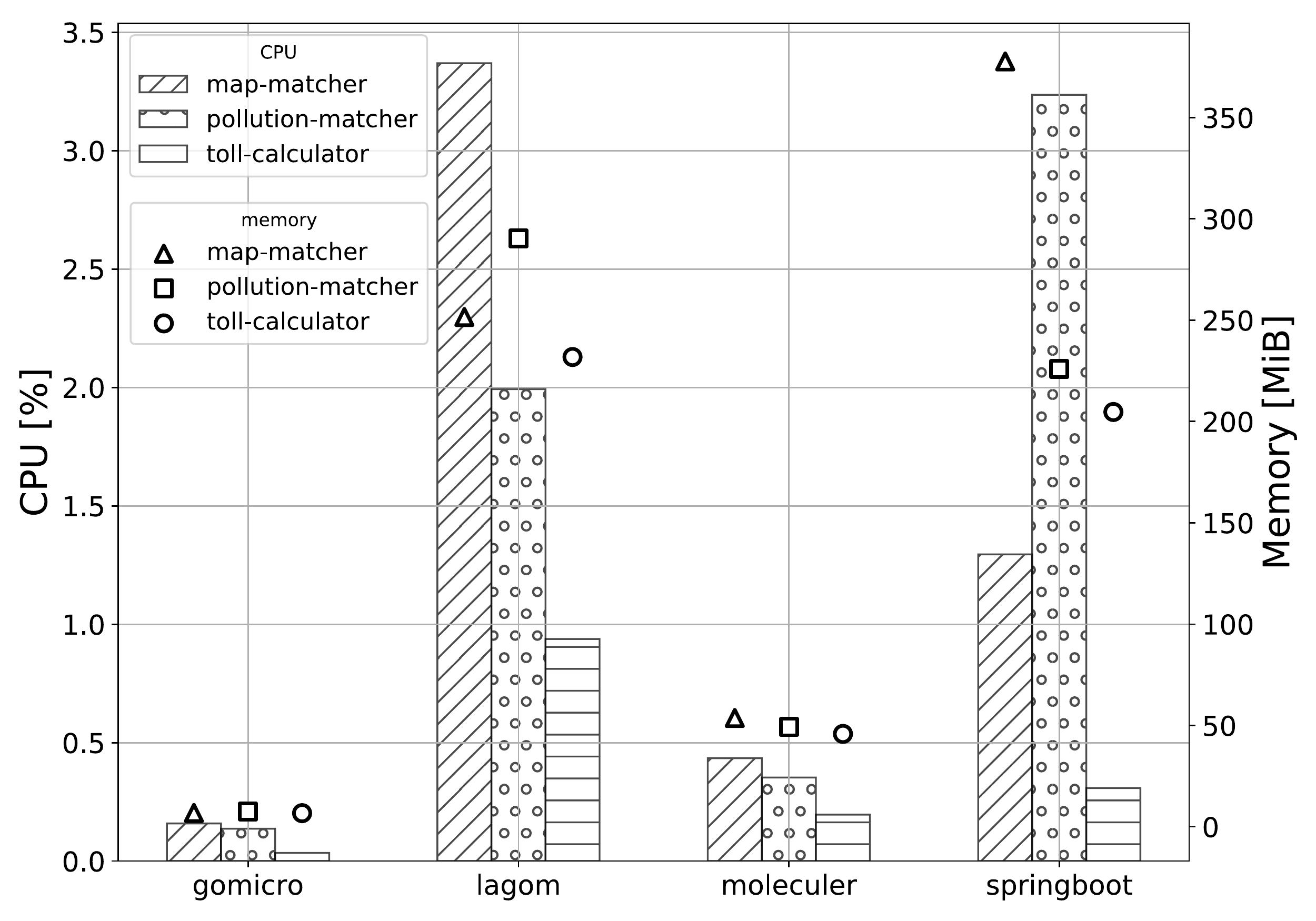}
	\caption{CPU and memory consumption of microservices implemented by different frameworks.}
	\label{fig:_memory}
\end{figure}

\paragraph{CPU and memory usage} As visible in Fig. \ref{fig:_memory}, Lagom and Spring yield much higher resource consumption than the other two frameworks. It is interesting that while Lagom has a higher CPU usage on the Map Matcher service than Spring, and lower than for the Pollution Matcher, in the case of memory usage the results are reversed. Looking again at Fig. \ref{fig:toll_performance}, the latency for the Map Matcher is the lowest with Spring, but this framework also shows the highest latency in the Pollution Matcher. 
Lagom obtains a slightly lower total end-to-end latency than Go Micro, but with a much higher resource consumption, so it is important to look at both aspects when deciding for a framework.

The lower CPU and memory consumption and small image size of the Go Micro services make it the most efficient framework in resource usage out of the four.

\section{Discussion} \label{discussion}
Three out of four frameworks studied in this work (Lagom, Moleculer, and Go Micro) encourage developers to structure the communication protocol before developing actual microservices. In contrast, Spring Boot doesn't force developers to follow a certain code style to define application's API. One of the possible reasons is that Spring Boot is a general-purpose framework, not designed specifically for microservices. This fact emphasizes the importance of communication pattern in distributed systems in general and in microservices in particular.

In theory, service decoupling is described as one of the most important features of microservices. However, in reality, there are several challenges to realize total independence among microservices. For example, shared libraries among microservices improve productivity of developers but also create dependencies. In our implementation, JSON is selected as the message format exchanged among services because it doesn't create specific requirements for microservices in serialization/deserialization. If Protocol Buffer is used instead, it forces every microservice to use ProtoBuf. However, given that ProtoBuf implementations are available for many programming languages, developers still have many options to develop their services. In another extreme end, using some platform-specific protocols such as \textit{Spring HTTP Invoker} forces both server and client to be written in a JVM language.

For a cloud native architecture such as microservices, interoperability is becoming more significant than ever. Adopting a microservice framework using standards or popular tools can be much more beneficial for an organization in the long run. In this regard, as one of the most popular software frameworks, Spring Boot/Spring Cloud provides developers a clear advantage to build microservices as this framework has an extensive level of support and integration available. Both Go Micro and Moleculer provide less opinionated approach to build microservices. Lagom relies heavily on Lightbend platform and Akka, so it might require additional efforts to deploy on other platforms.
 
 The development of microservices in particular, or cloud native applications in general, relies on many different levels of abstraction, which can degrade the performance significantly. Software architects and developers should also keep that in mind when selecting a technology stack for an application. For example, our results show that Moleculer can achieve a good end-to-end latency performance, but when deploying using Docker, Go Micro can produce much smaller images, which could be beneficial on limited resources. Similarly, JVM-based languages can lead to bigger Docker images, and higher resource consumption. Together with the JVM warm-up issue, it might not be the best choice for latency-sensitive application.
    
\section{Conclusion} \label{conclusion}
In this work, we presented an evaluation of microservice frameworks through the development of an example application, which includes both internal and external communication, data query from external database, and uses Docker for deployment. We defined three main comparison categories: For \textit{features}, we examined three categories, which are maturity, development supports, and built-in features; for \textit{design patterns}, we categorized them into six groups as presented in Table \ref{tab:design-table}; for \textit{performance}, we relied on two main criteria, which are end-to-end latency performance and resource consumption. The criteria mentioned in our evaluation can be extended to use as a comparison framework for selecting a development stack for a microservice-based application.

From the results presented in this work, it is clear that there is no one-size-fits-all solution when choosing a microservice framework. It depends a lot on application's specific functional and non-functional requirements. All of the frameworks mentioned in this paper as well as most of other options available can provide abstraction and boilerplate code to help developers deal with the complexities commonly found in distributed systems such as microservices. 
As shown in the evaluation section, interoperability is one of the most important criteria to maximize the flexibility in tool selection. For example, the fact that Lagom is designed to heavily rely on Akka and Lightbend platform, or Moleculer uses a custom naming convention for message queues, might cause integration issues with other microservices. Regarding design patterns, all frameworks follow asynchronous, event-driven architecture with supports for other capabilities such as distributed tracing or fault tolerance. When it comes to performance comparison, our results show how using JVM-based languages can result in heavier Docker images and higher resource consumption.

The design of our test is, however, still limited to provide a comprehensive comparison among selected frameworks. Although we have tried our best to have a similar implementation in different frameworks/languages, there are still some specific features that cannot be easily translated into other implementations. In addition, the interaction among microservices of our design is only a simple chain, which cannot convey more complicated interaction schemes. 

\section{Future work} \label{future}
As mentioned in the conclusions, an evaluation scenario with more complicated interaction patterns instead of chained microservices can produce more insights into the application performance. Other possibilities are to evaluate on the effects of data persistence strategies on microservices performance and scaling of the system, with multiple instances of each microservice where also load balancing could be tested.

\printbibliography

\end{document}